# Noncommutative Supersymmetric Yang-Mills Theory in Ten-Dimensions with Higher-Derivative Terms

Hitoshi Nishino[1] and Subhash Rajpoot[2]

*Department of Physics & Astronomy*
*California State University*
*Long Beach, CA 90840*

## Abstract

We present an action for noncommutative supersymmetric Yang-Mills theory in ten-dimensions, and confirm its invariance under supersymmetry. We next add higher-order derivative terms to such a noncommutative supersymmetric action. These terms contain fields as high as the quartic order. This resulting action can be regarded as supersymmetric generalization of noncommutative non-Abelian Dirac-Born-Infeld action. Some ambiguities related to field redefinitions are also clarified.

---

[1]E-Mail: hnishino@csulb.edu

[2]E-Mail: rajpoot@csulb.edu

## 1. Introduction

The idea that the space-time coordinates should be noncommutative [1] has been motivated by the development of open strings or D-branes, leading to the constant background antisymmetric field. Accordingly, the low energy effective theory of such open strings attached to noncommutative branes becomes a noncommutative gauge theory [2]. It has been also proven [3] that the noncommutative Dirac-Born-Infeld (DBI) theory is equivalent to the ordinary DBI theory under what is called Seiberg-Witten map [3][4]. The supersymmetrization of DBI theory with non-vanishing $B_{\mu\nu}$-field was studied in [5], and the conclusion was that it leads to the noncommutative supersymmetric gauge theory in a certain limit. A supersymmetric noncommutative DBI theory in 4D has been studied in [6] both for Abelian and non-Abelian gauge groups, based on superfield formulation. As in the non-supersymmetric case, the possible non-Abelian gauge group for noncommutative DBI theory is to be $U(N)$. However, this restriction on Yang-Mills gauge groups has been recently overcome by the use of anti-automorphism of $\star$ matrix algebra, consistently restricting the $u_\star(N)$ algebra to be $o_\star(N)$ or $usp_\star(N)$ algebras [7].

In this present Letter, we will present the noncommutative version of supersymmetric Yang-Mills theory in ten-dimensions (10D) with the gauge group $U(N)$, first for the covariant kinetic terms for a non-Abelian Yang-Mills multiplet, and next with its generalization with higher-derivative quartic terms that can be added to the kinetic terms containing the next-leading terms in the DBI action, up to quintic terms in fields. To put it differently, we will study the supersymmetrization [8][9][10] of noncommutative DBI theory [3][4] in 10D. As the guiding principle, we follow the result in [8][10] for commutative supersymmetric Yang-Mills theory with higher-order derivatives. We will show that by introducing a total symmetrization operator as in [6], the whole computation is drastically simplified, by avoiding the potentially dangerous ordering problem with $\star$ products at quartic order.

## 2. Noncommutative Supersymmetric Yang-Mills Theory in 10D

We start with the covariant kinetic terms of noncommutative supersymmetric Yang-Mills theory in 10D. Here we do not include higher-derivative terms, but consider only the kinetic terms. We first fix the total action $I_{\rm NCSYM}$ to be

$$I_{\rm NCSYM} = \int d^{10}x \ {\rm tr} \left[ +\tfrac{1}{4} F_{\mu\nu} \star F^{\mu\nu} + 2\overline{\lambda} \star \gamma^\mu D_\mu \lambda \right]$$
$$\equiv \int d^{10}x \left[ -\tfrac{1}{4} F_{\mu\nu}{}^I \star F^{\mu\nu\, I} - 2\overline{\lambda}^I \star \gamma^\mu D_\mu \lambda^I \right] \ , \quad (2.1)$$

where the tr-operator acts like ${\rm tr}\,(T^I T^J) = -\delta^{IJ}$ for anti-hermitian generators $T^I$ ($I, J, \cdots = 1, 2, \cdots, N^2$) of the gauge group $U(N)$ for the fields $A_\mu \equiv A_\mu{}^I T^I$ and $\lambda \equiv \lambda^I T^I$.

The $\star$ product is the noncommutative product defined by

$$f \star g \equiv f \, \exp\left(i\overleftarrow{\partial}_\mu \theta^{\mu\nu} \overrightarrow{\partial}_\nu\right) g$$
$$\equiv \sum_{n=0}^{\infty} \frac{i^n}{n!} \theta^{\mu_1\nu_1} \cdots \theta^{\mu_n\nu_n} (\partial_{\mu_1} \cdots \partial_{\mu_n} f)(\partial_{\nu_1} \cdots \partial_{\nu_n} g) \ . \quad (2.2)$$

The field strength $F_{\mu\nu}{}^I$ and the covariant derivative $D_\mu \lambda^I$ are defined by

$$F_{\mu\nu}{}^I \equiv \partial_\mu A_\nu{}^I - \partial_\nu A_\mu{}^I + (A_\mu \star A_\nu - A_\nu \star A_\mu)^I \equiv \partial_\mu A_\nu{}^I - \partial_\nu A_\mu{}^I + [A_\mu, A_\nu]_\star^I \ , \quad (2.3a)$$
$$D_\mu \lambda^I \equiv \partial_\mu \lambda^I + (A_\mu \star \lambda)^I - (\lambda \star A_\mu)^I \equiv \partial_\mu \lambda^I + [A_\mu, \lambda]_\star^I \ , \quad (2.3b)$$

as has been given by many authors [1][3][6]. The reason we put the factor $2$ in the $\lambda$-kinetic term is in order to comply with the notation in [8]. Relevantly, under $U(N)$ the fields transform as

$$\delta_\alpha A_\mu{}^I = \partial_\mu \alpha^I + [A_\mu, \alpha]_\star^I \ , \quad (2.4a)$$
$$\delta_\alpha \lambda^I = -[\alpha, \lambda]_\star^I \ . \quad (2.4b)$$

To save space, we sometimes omit the indices $I, J, \cdots$, using also the $\star$ commutator $[A, B]_\star \equiv A \star B - B \star A$. We skip the details for the confirmation of gauge invariance of our action $I_\text{NCSYM}$, due to its common feature shared with other 4D cases [1][3].

Our action $I_\text{NCSYM}$ in (2.1) is also invariant under supersymmetry

$$\delta_Q A_\mu{}^I = -(\overline{\epsilon} \gamma_\mu \lambda^I) \ , \quad (2.5a)$$
$$\delta_Q \lambda^I = +\tfrac{1}{8} \gamma^{\mu\nu} F_{\mu\nu}{}^I \ . \quad (2.5b)$$

The superinvariance of $I_\text{NCSYM}$ is confirmed by the frequent use of basic relations, such as

$$\int d^{10}x \, f \star g = \int d^{10}x \, g \star f = \int d^{10}x \, fg \ , \quad (2.6)$$

namely, the $\star$ product of two fields does not matter under the 10D integral $\int d^{10}x$, because the difference is only a total divergence. In the variation of the kinetic term of $A_\mu$, a convenient lemma is

$$\int d^{10}x \left( A \star [B, C]_\star - [A, B]_\star \star C \right) \equiv 0 \ , \quad (2.7)$$

for arbitrary fields $A$, $B$ and $C$, formulated as a corollary of (2.6). Another useful corollary is about the partial integration for the covariant derivative $D_\mu$:

$$\int d^{10}x \, B^I \star D_\mu C^I = -\int d^{10}x \, (D_\mu B^I) \star C^I \ , \quad (2.8)$$

which is used in the variation of the $\lambda$-kinetic term. Needless to say, the Bianchi identity

$$D_{[\mu}F_{\nu\rho]}{}^I \equiv 0 \ , \tag{2.9}$$

plays an important role in the confirmation of superinvariance.

As in the commutative case, after the cancellation of all the quadratic terms, we are left with the cubic term of the $\lambda$-field:

$$\delta_Q I_{\text{NCSYM}} = \int d^{10}x \left[ -2f^{IJK}(\bar{\epsilon}\gamma_\mu\lambda^I) \star (\bar{\lambda}^J \star \gamma^\mu\lambda^K) \right] \ , \tag{2.10}$$

where $f^{IJK}$ is the totally antisymmetric structure constant for $U(N)$. As in the commutative case, we need the Fierz identity

$$(\bar{\epsilon} \star \psi_1) \star (\bar{\psi}_2 \star \psi_3)$$
$$= -\tfrac{1}{16}(\bar{\psi}_1 \star \psi_2) \star (\bar{\epsilon} \star \psi_3) - \tfrac{1}{16}(\bar{\psi}_1 \star \gamma_\mu\psi_2) \star (\bar{\epsilon} \star \gamma^\mu\psi_3)$$
$$+ \tfrac{1}{32}(\bar{\psi}_1 \star \gamma_{\mu\nu}\psi_2) \star (\bar{\epsilon} \star \gamma^{\mu\nu}\psi_3) + \tfrac{1}{96}(\bar{\psi}_1 \star \gamma_{\mu\nu\rho}\psi_2) \star (\bar{\epsilon} \star \gamma^{\mu\nu\rho}\psi_3)$$
$$- \tfrac{1}{384}(\bar{\psi}_1 \star \gamma_{\mu\nu\rho\sigma}\psi_2) \star (\bar{\epsilon} \star \gamma^{\mu\nu\rho\sigma}\psi_3) - \tfrac{1}{1920}(\bar{\psi}_1 \star \gamma_{\mu\nu\rho\sigma\tau}\psi_2) \star (\bar{\epsilon} \star \gamma^{\mu\nu\rho\sigma\tau}\psi_3) \ , \tag{2.11}$$

for arbitrary Majorana-Weyl spinors $\epsilon$, $\psi_1$, $\psi_2$ and $\psi_3$. Depending on their chiralities, some terms above vanish, e.g., $(\bar{\psi}_1 \star \psi_2) \equiv 0$ if $\psi_1$ and $\psi_2$ share the same chirality. The Fierzing (2.11) is possible without a total symmetrization to be used in the next section, thanks to the constancy of the parameter $\epsilon$. After applying the Fierzing (2.11) to (2.10), the question $\delta_Q I_{\text{NCSYM}} \stackrel{?}{=} 0$ is now equivalent to

$$0 \stackrel{?}{=} +\int d^{10}x \, f^{IJK}(\bar{\lambda}^I \star \gamma^{\rho\sigma\tau}\lambda^J) \star (\bar{\epsilon}\gamma_{\rho\sigma\tau}\lambda^K)$$
$$= +2(\gamma^{\rho\sigma\tau})_{\alpha[\beta|}(\gamma_{\rho\sigma\tau})_{|\gamma]\delta}\, \epsilon^\delta \int d^{10}x \ \text{tr}\, (\lambda^a \star \lambda^\beta \star \lambda^\gamma) \ . \tag{2.12}$$

This term with three-gamma matrix sandwiched by the $\lambda$'s was absent in the *commutative* case, because of the $(\alpha\beta)$ symmetry of $f^{IJK}\lambda^{I\alpha}\lambda^{J\beta}$. But now we have this because the latter is to be replaced by $f^{IJK}\lambda^{I\alpha} \star \lambda^{J\beta}$, which has no such symmetry in $(\alpha\beta)$ due to the $\star$ product. This problem is solved by the $\gamma$-matrix identity

$$(\gamma^{\rho\sigma\tau})_{\alpha[\beta|}(\gamma_{\rho\sigma\tau})_{|\gamma\delta]} \equiv 0 \ , \tag{2.13}$$

so that (2.12) equals

$$0 \stackrel{?}{=} +\int d^{10}x \, f^{IJK}(\bar{\lambda}^I \star \gamma^{\rho\sigma\tau}\lambda^J) \star (\bar{\epsilon}\gamma_{\rho\sigma\tau}\lambda^K)$$
$$= +2(\gamma^{\rho\sigma\tau})_{\alpha[\beta|}(\gamma_{\rho\sigma\tau})_{|\gamma]\delta}\, \epsilon^\delta \int d^{10}x \ \text{tr}\, (\lambda^a \star \lambda^\beta \star \lambda^\gamma)$$
$$= +(\gamma^{\rho\sigma\tau})_{\alpha\beta}(\gamma_{\rho\sigma\tau})_{\gamma\delta}\, \epsilon^\delta \int d^{10}x \ \text{tr}\, (\lambda^a \star \lambda^\beta \star \lambda^\gamma)$$
$$= +\tfrac{1}{2}\int d^{10}x \, f^{IJK}(\bar{\lambda}^I \star \gamma^{\rho\sigma\tau}\lambda^J) \star (\bar{\epsilon}\gamma_{\rho\sigma\tau}\lambda^K) \ , \tag{2.14}$$

proportional to the first line with the constant $1/2$. This implies that the first line itself is to vanish. The important ingredient here is that even though there is a three-gamma matrix sandwiched by the $\lambda$'s, such a term vanishes by itself due to the $\gamma$-algebra (2.13). This concludes the confirmation of superinvariance of the action $I_{\rm NCSYM}$.

## 3. Noncommutative Supersymmetric Theory with Higher-Derivatives in 10D

We next consider the inclusion of higher-order derivatives into the supersymmetric Abelian gauge theory in 10D. Since the resulting lagrangian will contain the bosonic noncommutative DBI lagrangian, we can regard this also as the supersymmetrization of noncommutative DBI theory. In this sense, we call our action noncommutative supersymmetric DBI action represented by $I_{\rm NCSDBI}$.

In this paper we fix our lagrangian for $I_{\rm NCSDBI}$ with all the quartic terms at $\mathcal{O}(\alpha^2)$ including also $\lambda^4$-terms. However, the terms quintic in fields are ignored at $\mathcal{O}(\alpha^2)$, just for simplicity of computation. These quintic terms had been also ignored in [8], but has been analyzed recently in superspace formulation [10]. These quintic terms arise only in the commutators of gauge-covariant derivatives that always contains the structure constant of the gauge group [10]. In this paper, these quintic terms are ignored just for simplicity, such that partial integrations can be done rather easily. As for the transformation rule for the Yang-Mills field $A_\mu$ and the gaugino $\lambda$, we fix terms only up to $\lambda^3$ and $\lambda^2 F$-terms, respectively.

We first summarize our result here. Our total action $I_{\rm NCSDBI}$ is

$$I_{\rm NCSDBI} \equiv \int d^{10}x\, \mathcal{L}_{\rm NCSDBI} \ , \tag{3.1}$$

$$\begin{aligned}\mathcal{L}_{\rm NCSDBI} \equiv &-\tfrac{1}{4} F_{\mu\nu}{}^I \star F^{\mu\nu I} - 2(\overline{\lambda}^I \star \gamma^\mu D_\mu \lambda^I) \\ &+ \alpha^2\, {\rm tr}\, \mathcal{S}^\star \Big[ -\tfrac{1}{4}(F\star F\star F\star F)_\mu{}^\mu + \tfrac{1}{16}(F\star F)_\mu{}^\mu \star (F\star F)_\nu{}^\nu \\ &\quad + 2(F\star F)^{\mu\nu} \star (\overline{\lambda} \star \gamma_\mu D_\nu \lambda) + \tfrac{1}{2} F_\mu{}^\lambda \star (D_\lambda F_{\nu\rho}) \star (\overline{\lambda}\star \gamma^{\mu\nu\rho}\lambda) \\ &\quad - \tfrac{4}{3}(\overline{\lambda}\star \gamma_\mu D_\nu \lambda)\star(\overline{\lambda}\star \gamma^\mu D^\nu \lambda) \Big] + \mathcal{O}(\alpha^2 \varphi^5) + \mathcal{O}(\alpha^3)\ . \end{aligned} \tag{3.2}$$

Here $\alpha$ is a constant with the dimension of (length)$^2$ in order to keep track of the higher-derivative terms. For example, we sometimes call the first line in (3.2) $\mathcal{L}_{\alpha^0}$, while the rest $\mathcal{L}_{\alpha^2}$. The symbol $\mathcal{O}(\alpha^2 \varphi^5)$ represents any terms quintic in fields at $\mathcal{O}(\alpha^2)$, while $\mathcal{O}(\alpha^3)$ is for terms at order higher than $\mathcal{O}(\alpha^2)$.

The operator $\mathcal{S}^\star$ is for the total symmetrization operator, defined by

$$\mathcal{S}^\star(A \star B \star C \star D)$$
$$\equiv \tfrac{1}{4}\Big[ A \star \mathcal{S}^\star(B \star C \star D) + (-1)^{A(B+C+D)} B \star \mathcal{S}^\star(C \star D \star A)$$
$$+ (-1)^{(C+D)(A+B)} C \star \mathcal{S}^\star(D \star A \star B) + (-1)^{D(A+B+C)} D \star \mathcal{S}^\star(A \star B \star C) \Big] \ , \quad (3.3\text{a})$$
$$\mathcal{S}^\star(A \star B \star C)$$
$$\equiv \tfrac{1}{6}\Big[ A \star B \star C + (-1)^{A(B+C)} B \star C \star A + (-1)^{C(A+B)} C \star A \star B$$
$$+ (-1)^{CB} A \star C \star B + (-1)^{AB} B \star A \star C + (-1)^{C(A+B)+AB} C \star B \star A \Big] \ , \quad (3.3\text{b})$$

where the superscripts $A$, $B$, $C$, $D$ are for the Grassmann parities of each field in $A$, $B$, $C$, $D$. Effectively, under the $\mathcal{S}^\star$-operations, the ordering problem with the $\star$ product disappears [6], because of the total symmetrization by definition. Note also that the tr-operation acts on the anti-hermitian generators $T^I$, after the total symmetrization by the $\mathcal{S}^\star$'s.

The symbols such as $(F \star F \star F \star F)_\mu{}^\nu$ are defined by

$$(F \star F)_\mu{}^\nu \equiv F_\mu{}^\rho \star F_\rho{}^\nu \ , \quad (F \star F \star F \star F)_\mu{}^\nu \equiv F_\mu{}^\rho \star F_\rho{}^\sigma \star F_\sigma{}^\tau \star F_\tau{}^\nu \ . \quad (3.4)$$

Compared with the commutative case [8][10], all the coefficients in (3.2) are in agreement with the commutative case, *except for* the special usage of the total symmetrization $\mathcal{S}^\star$ for $\star$ products. If we look only at purely bosonic terms, they are of the form

$$+ \tfrac{1}{4} \operatorname{tr}(F \star F)_\mu{}^\mu - \tfrac{1}{4}\alpha^2 \operatorname{tr} \mathcal{S}^\star\Big[ (F \star F \star F \star F)_\mu{}^\mu \Big] \ , \quad (3.5)$$

which are the terms at $\mathcal{O}(\alpha^0)$ and $\mathcal{O}(\alpha^2)$ in the noncommutative DBI lagrangian [6][3]

$$\mathcal{L}_{DBI} \equiv b^{-2}\alpha^{-2} \operatorname{tr} \mathcal{S}^\star \left[ \sqrt[\star]{\det{}^\star(\delta_\mu{}^\nu + b\alpha F_\mu{}^\nu)} \right] \ . \quad (3.6)$$

Therefore, our action $I_{\text{NCSDBI}}$ can also be regarded as the supersymmetrization of that of the noncommutative DBI theory.

Our $I_{\text{NCSDBI}}$ is invariant under supersymmetry

$$\delta_Q A_\mu = -(\bar{\epsilon}\gamma_\mu \lambda)$$
$$+ \alpha^2 \mathcal{S}^\star \Big[ + \tfrac{3}{8}(F \star F)_\nu{}^\nu \star (\bar{\epsilon}\gamma_\mu \lambda) - (F \star F)_\mu{}^\nu \star (\bar{\epsilon}\gamma_\nu \lambda)$$
$$- \tfrac{1}{4} F_{\mu\nu} \star F_{\rho\sigma} \star (\bar{\epsilon}\gamma^{\nu\rho\sigma}\lambda) + \tfrac{1}{16} F^{\rho\sigma} \star F^{\tau\lambda} \star (\bar{\epsilon}\gamma_{\mu\rho\sigma\tau\lambda}\lambda) \Big] + \mathcal{O}(\alpha^2 \lambda^3) \ , \quad (3.7\text{a})$$
$$\delta_Q \lambda = + \tfrac{1}{8}(\gamma^{\mu\nu}\epsilon) F_{\mu\nu}$$
$$+ \alpha^2 \mathcal{S}^\star \Big[ + \tfrac{1}{64}(F \star F)_\nu{}^\nu \star F_{\sigma\tau}(\gamma^{\sigma\tau}\epsilon) - \tfrac{1}{16}(F \star F \star F)_{\mu\nu}(\gamma^{\mu\nu}\epsilon)$$
$$- \tfrac{1}{384} F_{\mu\nu} \star F_{\rho\sigma} \star F_{\tau\lambda} \star (\gamma^{\mu\nu\rho\sigma\tau\lambda}\lambda) \Big] + \mathcal{O}(\alpha^2 \lambda^2 F) \ , \quad (3.7\text{a})$$

---

[3]The noncommutative square root $\sqrt[\star]{1+x}$ is defined by the expansion $\sqrt[\star]{1+x} \equiv 1 + \sum_{n=1}^\infty (1/n!)(1/2)(1/2-1)\cdots(3/2-n) \overbrace{x \star x \star \cdots \star x}^{n}$ [6], while the $\det^\star$ is the noncommutative determinant.

where the adjoint indices are suppressed, but are taken for granted.

The confirmation of the supersymmetric invariance of our action $I_{\text{NCSDBI}}$ up to $\mathcal{O}(\alpha^2 \varphi^5)$ and $\mathcal{O}(\alpha^3)$-terms is performed as follows. First, note that there arise two sorts of terms, when we vary $I_{\text{NCSDBI}}$: (I) $F^3\lambda$-terms, and (II) $F\lambda^3$-terms. Next, we take care of these two categories in turn:

As for the (I) $\alpha^2 F^3\lambda$-terms, there are three sources of these terms: (i) $\delta_Q \mathcal{L}_{\alpha^2 F^4}|_{\alpha^2 F^3 \lambda}$, (ii) $\delta_Q \mathcal{L}_{\alpha^2 F^2 \lambda^2}|_{\alpha^2 F^3 \lambda}$, and (iii) $\delta_Q \mathcal{L}_{\alpha^0}|_{\alpha^2 F^3 \lambda}$.

We start with the sectors (i) and (ii). To this end, we first establish the convenient lemma for the variation of $\delta_Q A_\mu$:

$$\alpha^2 \mathcal{S}^\star \Big[ \{\delta_Q (F \star F)_{\mu\nu}\} \star X^{\mu\nu} \Big]$$
$$= \mathcal{S}^\star \Big[ 2\alpha^2 (\bar{\epsilon} \gamma^\rho \lambda) \star D_\mu (F_{\rho\nu} \star X^{\mu\nu}) - 2\alpha^2 (\bar{\epsilon} \gamma_\mu \lambda) \star D_\rho (F^\rho{}_\nu \star X^{\mu\nu}) \Big] + \mathcal{O}(\alpha^2 \varphi^4) \ , \quad (3.8)$$

where $X^{\mu\nu}$ is an arbitrary field or $\star$ products of fields. Eq. (3.8) is up to a total divergence in 10D, as well as quartic terms in fields at $\mathcal{O}(\alpha^2)$. By the aid of this lemma, we can arrange all the contributions in (i) and (ii) as

$$\delta_Q \mathcal{L}_{\alpha^2}|_{\alpha^2 \lambda F^3}$$
$$= \alpha^2 \operatorname{tr} \mathcal{S}^\star \Big[ + \tfrac{1}{2}(\bar{\epsilon} \gamma^\rho \lambda) \star (F \star F)^{\lambda\nu} \star D_\lambda F_{\nu\rho} + \tfrac{3}{2}(\bar{\epsilon} \gamma^\rho \lambda) \star (F \star F)_{\rho\tau} \star D_\nu F^{\nu\tau}$$
$$- \tfrac{1}{2}(\bar{\epsilon} \gamma_\rho \lambda) \star (F \star F)_\sigma{}^\sigma \star D_\mu F^{\mu\rho} + \tfrac{1}{4}(\bar{\epsilon} \gamma_\mu{}^{\rho\sigma} \lambda) \star (F \star F)^{\mu\nu} D_\nu F_{\rho\sigma}$$
$$- \tfrac{1}{8}(\bar{\epsilon} \gamma^{\mu\rho\sigma} \lambda) \star F_{\rho\sigma} \star F_{\tau\nu} \star D_\mu F^{\tau\nu} - \tfrac{1}{4}(\bar{\epsilon} \gamma^{\mu\rho\sigma} \lambda) \star F_{\mu\tau} \star F_{\rho\sigma} \star D_\nu F^{\nu\tau}$$
$$+ \tfrac{1}{2}(\bar{\epsilon} \gamma^{\mu\sigma\nu} \lambda) \star F_\mu{}^\lambda \star F_{\sigma\rho} \star D_\lambda F_\nu{}^\rho - \tfrac{1}{8}(\bar{\epsilon} \gamma^{\sigma\tau\mu\nu\rho} \lambda) \star F_\mu{}^\lambda \star F_{\sigma\tau} \star D_\lambda F_{\nu\rho} \Big] \ . \quad (3.9)$$

The guiding principle for arranging these terms is that any derivative hitting on the $\lambda$-field should be partially integrated, such that only one of the $F$'s in the three $F$'s should be hit by a derivative. We can always perform such a partial integration, yielding only those terms in (3.9). There was also an exact cancellation between the like terms of the type $(\bar{\epsilon} \gamma^\rho \lambda) \star F_{\nu\rho} \star F_{\mu\sigma} \star D^\nu F^{\mu\sigma}$, which is thus absent in (3.9).

We next look into the sector (iii). For this sector, we need a special lemma related to the $\mathcal{S}^\star$-operation. This is because when we vary the kinetic terms at $\mathcal{O}(\alpha^0)$, we need to substitute the $\mathcal{O}(\alpha^2)$-terms in (3.7). However, the $\mathcal{S}^\star$-operation in such terms in (3.7) symmetrizes only three fields, while these terms from this sector (iii) are supposed to cancel those terms in (3.9), where all the $\mathcal{S}^\star$-operations symmetrize all the four fields. A convenient lemma to solve this problem is

$$\int d^{10}x \ \operatorname{tr} \Big[ \mathcal{S}^\star (A \star B \star C) \star D \Big] \equiv \int d^{10}x \ \operatorname{tr} \mathcal{S}^\star (A \star B \star C \star D) \ . \quad (3.10)$$

This lemma means that any $\star$ product of a $\mathcal{S}^\star$-symmetrized three fields with a field equals the total symmetrization of the $\star$ product of all the four fields. This lemma is easily confirmed by the use of the fundamental identity, such as (2.6), considering also the Grassmann parities of all the fields. In fact, the l.h.s. of (3.10) is

$$\begin{aligned}(\text{r.h.s.}) &= \int d^{10}x \tfrac{1}{24}\Big[ + A \star B \star C \star D + (-1)^{A(B+C)} B \star C \star A \star D + (-1)^{C(A+B)} C \star A \star B \star D \\ &\quad + (-1)^{AB} B \star A \star C \star D + (-1)^{A(B+C)+BC} C \star B \star A \star D + (-1)^{BC} A \star C \star B \star D \\ &\quad + (-1)^{A(B+C+D)} B \star C \star D \star A + (-1)^{A(B+C+D)+B(C+D)} C \star D \star B \star A \\ &\quad + (-1)^{A(B+C+D)+D(B+C)} D \star B \star C \star A \\ &\quad + (-1)^{A(B+C+D)+BC} C \star B \star D \star A + (-1)^{A(B+C+D)+B(C+D)+CD} D \star C \star B \star A \\ &\quad + (-1)^{A(B+C+D)+CD} B \star D \star C \star A \\ &\quad + (-1)^{(A+B)(C+D)} C \star D \star A \star B + (-1)^{(A+B)(C+D)+C(D+A)} D \star A \star C \star B \\ &\quad + (-1)^{B(C+D)} A \star C \star D \star B \\ &\quad + (-1)^{(A+B)(C+D)+CD} D \star C \star A \star B + (-1)^{B(C+D)+CD} A \star D \star C \star B \\ &\quad + (-1)^{B(C+D)+AC} C \star A \star D \star B \\ &\quad + (-1)^{D(A+B+C)} D \star A \star B \star C + (-1)^{CD} A \star B \star D \star C \\ &\quad + (-1)^{CD+A(B+D)} B \star D \star A \star C \\ &\quad + (-1)^{D(B+C)} A \star D \star B \star C + (-1)^{AB+CD} B \star A \star D \star C \\ &\quad + (-1)^{(B+C)D+A(B+D)} D \star B \star A \star C \Big] \; , \end{aligned} \qquad (3.11)$$

where $(-1)^A$ represent the usual Grassmann parity of the $A$-field, namely, $(-1)^0 = +1$ when the field $A$ is bosonic, while it is $(-1)^{+1} = -1$ when fermionic. Next we use the property (2.6) for the r.h.s. of (3.11), such that the field $D$ is always at the end:

$$\begin{aligned}(\text{r.h.s.}) &= \int d^{10}x \tfrac{1}{24}\Big[ + 6\mathcal{S}^\star(A \star B \star C) \star D \\ &\quad + A \star B \star C \star D + (-1)^{BA} B \star A \star C \star D + (-1)^{A(B+C)} B \star C \star A \star D \\ &\quad + (-1)^{BC} A \star C \star B \star D + (-1)^{A(B+C)+BC} C \star B \star A \star D + (-1)^{C(A+B)} C \star A \star B \star D \\ &\quad + A \star B \star C \star D + (-1)^{BC} A \star C \star B \star D + (-1)^{BA} B \star A \star C \star D \\ &\quad + (-1)^{C(A+B)} C \star A \star B \star D + (-1)^{A(B+C)+BC} C \star B \star A \star D + (-1)^{A(B+C)} B \star C \star A \star D \\ &\quad + A \star B \star C \star D + (-1)^{C(A+B)} C \star A \star B \star D + (-1)^{BC} A \star C \star B \star D \end{aligned}$$

$$+ (-1)^{A(B+C)} B \star C \star A \star D + (-1)^{A(C+B)+BC} C \star B \star A \star D$$
$$+ (-1)^{AB} B \star A \star C \star D \Big] \quad , \tag{3.12}$$

Now the second and third lines in (3.12) are combined to yield $\mathcal{S}^\star(A \star B \star C) \star D$. The same is also true for the fourth and fifth lines of (3.12), and the remaining lines. After all, we get from (3.12) that the r.h.s. of (3.10) is

$$\begin{aligned}(\text{r.h.s.}) &= \int d^{10}x \, \tfrac{1}{24} \Big[ + 6\mathcal{S}^\star(A \star B \star C) \star D + 6\mathcal{S}^\star(A \star B \star C) \star D \\ &\qquad + 6\mathcal{S}^\star(A \star B \star D) + 6\mathcal{S}^\star(A \star B \star C) \star D \Big] \\ &= \int d^{10}x \, \mathcal{S}^\star(A \star B \star C) \star D = (\text{l.h.s.}) \quad .\end{aligned} \tag{3.13}$$

agreeing with the l.h.s. of (3.10).

After using the lemma (3.10), we can arrange all the contributions from (iii) $\delta_Q \mathcal{L}_{\alpha^0}|_{\alpha^2 F^3 \lambda}$ in such a way that all of them cancel exactly all the terms in (3.9). In other words, we get $\delta_Q \mathcal{L}_{\text{NCSDBI}}|_{\alpha^2 \lambda F^3} = \delta_Q (\mathcal{L}_{\alpha^2} + \mathcal{L}_{\alpha^0})|_{\alpha^2 \lambda F^3} = 0$ up to a total divergence.

As for the (II) $\alpha^2 F \lambda^3$-terms, we need to proceed with special care. As has been mentioned, we do not fix the $\mathcal{O}(\alpha^2 \lambda^3)$-terms in $\delta_Q A_\mu$ or $\mathcal{O}(\alpha^2 \lambda^2 F)$-terms in $\delta_Q \lambda$ in this paper. Consequently, any variation of the $\mathcal{O}(\alpha^0)$ kinetic terms that potentially contribute to the $\alpha^2 F \lambda^3$-terms *via* these terms in $\delta_Q A_\mu$ or $\delta_Q \lambda$ will not concern us here. Accordingly, we can also ignore terms with the factor $D_\nu F^{\mu\nu}$ and/or $\slashed{D}\lambda$ which can be absorbed into the modification of $\delta_Q A_\mu$ and $\delta_Q \lambda$, *via* the variation of the $\mathcal{O}(\alpha^0)$ kinetic terms. Considering these points, there are only two sources for this sector: (i) $\delta_Q \mathcal{L}_{\alpha^2 \lambda^4}|_{\alpha^2 F \lambda^3}$ and (ii) $\delta_Q \mathcal{L}_{\alpha^2 F^2 \lambda^2}|_{\alpha^2 F \lambda^3}$. Let us now consider the sectors (i) and (ii) in turn.

A simple consideration reveals, the sector (i) can produce only the terms of the type $\mathcal{S}^\star[(DF) \star (\bar{\epsilon}\gamma D\lambda) \star (\bar{\lambda}\gamma_{[3]}\lambda)]$, after appropriate partial integrations, where $\gamma$ in $(\bar{\epsilon}\gamma D\lambda)$ can be any odd number of $\gamma$'s. This is due to the chiralities of these fermionic fields, and the fact that under the $\mathcal{S}^\star$-operation, only three-gamma can be sandwiched by the two $\lambda$'s at the end. A further consideration shows that there are only five categories of such terms defined by

$$(1A) \equiv \alpha^2 \, \text{tr} \, \mathcal{S}^\star \Big[ (D_\nu F^{\rho\sigma}) \star (\bar{\epsilon}\gamma^\tau D^\nu \lambda) \star (\bar{\lambda} \star \gamma_{\rho\sigma\tau} \lambda) \Big] \quad , \tag{3.14a}$$

$$(3A) \equiv \alpha^2 \, \text{tr} \, \mathcal{S}^\star \Big[ (D_\nu F_{\rho\sigma}) \star (\bar{\epsilon}\gamma^{\rho\tau\lambda} D^\nu \lambda) \star (\bar{\lambda} \star \gamma^\sigma{}_{\tau\lambda} \lambda) \Big] \quad , \tag{3.14b}$$

$$(3B) \equiv \alpha^2 \, \text{tr} \, \mathcal{S}^\star \Big[ (D_\nu F_\mu{}^\rho) \star (\bar{\epsilon}\gamma^{\mu\tau\lambda} D_\rho \lambda) \star (\bar{\lambda} \star \gamma_{\tau\lambda}{}^\nu \lambda) \Big] \quad , \tag{3.14c}$$

$$(3C) \equiv \alpha^2 \, \text{tr} \, \mathcal{S}^\star \Big[ (D_\mu F_{\nu\rho}) \star (\bar{\epsilon}\gamma^{\mu\tau\lambda} D^\rho \lambda) \star (\bar{\lambda} \star \gamma_{\tau\lambda}{}^\mu \lambda) \Big] \quad , \tag{3.14d}$$

$$(5A) \equiv \alpha^2 \, \text{tr} \, \mathcal{S}^\star \Big[ (D_\mu F_{\rho\sigma}) \star (\bar{\epsilon}\gamma^{\rho\sigma\tau\lambda\omega} D^\mu \lambda) \star (\bar{\lambda} \star \gamma_{\tau\lambda\omega} \lambda) \Big] \quad . \tag{3.14e}$$

We can further see that these five terms are not really independent of each other. There is a relationship up to a term proportional to $D_\nu F^{\mu\nu}$ and $\displaystyle{\not}D\lambda$ for the reason already mentioned:

$$(5A) - 12(3A) - 42(1A) \doteq \mathcal{O}(\alpha^2) \ . \tag{3.15}$$

where $\doteq$ stands for an equality up to the factor of $D_\nu F^{\mu\nu} \doteq \mathcal{O}(\alpha^2)$ or $\displaystyle{\not}D\lambda \doteq \mathcal{O}(\alpha^2)$. Eq. (3.15) can be confirmed by the aid of Fierz identity for arbitrary Majorana-Weyl spinors $\psi_1, \cdots, \psi_4$:

$$\begin{aligned}
&\mathcal{S}^\star\big[(\overline{\psi}_1 \star \psi_2) \star (\overline{\psi}_3 \star \psi_4)\big] \\
&= \mathcal{S}^\star\Big[-\tfrac{1}{16}(\overline{\psi}_1 \star \psi_4) \star (\overline{\psi}_3 \star \psi_2) - \tfrac{1}{16}(\overline{\psi}_1 \star \gamma_\mu \psi_4) \star (\overline{\psi}_3 \star \gamma^\mu \psi_2) \\
&\quad + \tfrac{1}{32}(\overline{\psi}_1 \star \gamma_{\mu\nu} \psi_4) \star (\overline{\psi}_3 \star \gamma^{\mu\nu} \psi_2) + \tfrac{1}{96}(\overline{\psi}_1 \star \gamma_{\mu\nu\rho} \psi_4) \star (\overline{\psi}_3 \star \gamma^{\mu\nu\rho} \psi_2) \\
&\quad - \tfrac{1}{384}(\overline{\psi}_1 \star \gamma_{\mu\nu\rho\sigma} \psi_4) \star (\overline{\psi}_3 \star \gamma^{\mu\nu\rho\sigma} \psi_2) - \tfrac{1}{1920}(\overline{\psi}_1 \star \gamma_{\mu\nu\rho\sigma\tau} \psi_4) \star (\overline{\psi}_3 \star \gamma^{\mu\nu\rho\sigma\tau} \psi_2)\Big] \ ,
\end{aligned} \tag{3.16}$$

which is a totally symmetrized version of (2.11), now with the space-time dependent $\psi$'s which necessitates the $\mathcal{S}^\star$-operation. In order to get the relationship (3.15), we consider and simplify the following term

$$\begin{aligned}
&\text{tr}\, \mathcal{S}^\star\big[(D_\lambda F_{\nu\rho}) \star (\overline{\epsilon}\gamma_\mu \lambda) \star (\overline{\lambda} \star \gamma^{\mu\nu\rho} D^\lambda \lambda)\big] \\
&= \text{tr}\, \mathcal{S}^\star\big[+\tfrac{1}{2}(D_\lambda F_{\nu\rho}) \star (\overline{\epsilon}\gamma_\mu \lambda) \star D^\lambda(\overline{\lambda} \star \gamma^{\mu\nu\rho} \lambda)\big] \\
&\doteq \text{tr}\, \mathcal{S}^\star\big[-\tfrac{1}{2}(D_\lambda F_{\nu\rho}) \star (\overline{\epsilon}\gamma_\mu D^\lambda \lambda) \star (\overline{\lambda} \star \gamma^{\mu\nu\rho} \lambda)\big] = -\tfrac{1}{2}(1A) \ .
\end{aligned} \tag{3.17}$$

On the other hand, by the use of the Fierzing (3.16) applied to the case of $\overline{\psi}_1 \equiv \overline{\epsilon}\gamma_\mu$, $\psi_2 \equiv \lambda$, $\overline{\psi}_3 \equiv \overline{\lambda}$, $\psi_4 \equiv \gamma_{\mu\nu\rho} D_\lambda \lambda$, we get

$$\begin{aligned}
&\text{tr}\, \mathcal{S}^\star\big[(D_\lambda F_{\nu\rho}) \star (\overline{\epsilon}\gamma_\mu \lambda) \star (\overline{\lambda} \star \gamma^{\mu\nu\rho} D^\lambda \lambda)\big] \\
&= \text{tr}\, \mathcal{S}^\star\big[+\tfrac{1}{96}(D^\lambda F^\nu{}_\rho) \star (\overline{\epsilon}\gamma^\mu \gamma^{\sigma\tau\omega} \gamma_{\mu\nu\rho} D^\lambda \lambda) \star (\overline{\lambda} \star \gamma_{\sigma\tau\omega} \lambda)\big] \\
&= -\tfrac{1}{48}(5A) + \tfrac{1}{4}(3A) + \tfrac{3}{8}(1A) \ .
\end{aligned} \tag{3.18}$$

Equating (3.17) with (3.18), we get (3.15).

By the frequent use of these relations, we get the contribution from the sector (i) as

$$\delta_Q \mathcal{L}_{\alpha^2 \lambda^2}\big|_{\alpha^2 F\lambda^3} = \alpha^2\Big[+\tfrac{1}{36}(5A) + \tfrac{1}{6}(3A) - \tfrac{1}{6}(1A)\Big] \ , \tag{3.19}$$

while for the contribution from the sector (ii) as

$$\delta_Q \mathcal{L}_{\alpha^2 F^2 \lambda^2}\big|_{\alpha^2 F\lambda^3} = -\tfrac{1}{24}\alpha^2(5A) + \tfrac{3}{4}\alpha^2(1A) \ . \tag{3.20}$$

To reach this result, we also used the fact that

$$\alpha^2\, \text{tr}\, \mathcal{S}^\star\big[F^{\rho\nu} \star (\overline{\epsilon}\gamma_\mu \lambda) \star (D_\rho \overline{\lambda}) \star \gamma^\mu D_\nu \lambda\big] \doteq 0 \ , \tag{3.21}$$

as is easily confirmed by a Fierzing, up to the terms vanishing upon $D_\nu F^{\mu\nu} \doteq \mathcal{O}(\alpha^2)$ or $\rlap{/}{D}\lambda \doteq \mathcal{O}(\alpha^2)$, as well as up to a total divergence. Now, by the use of (3.15), we see that the two contributions from (3.19) and (3.20) exactly cancel each other. This concludes the summary of the invariance check of our total action $I_{\text{NCSDBI}}$ up to $\mathcal{O}(\alpha^2\varphi^5)$ and $\mathcal{O}(\alpha^3)$-terms.

## 4. Ambiguities Associated with Field Redefinitions

It is well-known in commutative supersymmetric gauge theory in 10D that there is some ambiguity about the coefficients of certain lagrangian terms caused by the freedom of field redefinitions [8][10]. This is also true with our noncommutative action $I_{\text{NCSDBI}}$. To see this, we consider the field redefinitions of the type

$$A_\mu \equiv A'_\mu - \tfrac{1}{4} c_2 \alpha^2 \mathcal{S}^\star \big[ F_{\nu\rho} \star (\overline{\lambda} \star \gamma_\mu{}^{\nu\rho}\lambda) \big] \ , \tag{4.1a}$$

$$\lambda \equiv \lambda' + \tfrac{1}{16} \alpha^2 \mathcal{S}^\star \big[ c_1 (F \star F)_\mu{}^\mu \star \lambda + c_3 F_{\mu\nu} \star F_{\rho\sigma} \star \gamma^{\mu\nu\rho\sigma}\lambda \big] \ , \tag{4.1b}$$

with arbitrary real constants $c_1$, $c_2$ and $c_3$. If we substitute (4.1) into our lagrangian (2.2), there arise some new terms with $c_1$, $c_2$, $c_3$ up to $\mathcal{O}(\alpha^4)$-terms, which we call $\Delta\mathcal{L}_{\text{NCSDBI}}$. The explicit form of $\mathcal{L}_{\text{NCSDBI}}$ is easily computed to be

$$\Delta\mathcal{L}_{\text{NCSDBI}} = \alpha^2 \, \text{tr}\, \mathcal{S}^\star \Big[ -\tfrac{1}{4} c_2 F_{\mu\nu} \star (D_\lambda F^\lambda{}_\rho) \star (\overline{\lambda} \star \gamma^{\mu\nu\rho}\lambda)$$
$$- \tfrac{1}{4} c_1 (F \star F)_\mu{}^\mu \star (\overline{\lambda}\rlap{/}{D}\lambda) - \tfrac{1}{4} c_3 F_{\mu\nu} \star F_{\rho\sigma} \star (\overline{\epsilon}\gamma^{\mu\nu\rho\sigma}\rlap{/}{D}\lambda) \Big] \ . \tag{4.2}$$

For the reason already mentioned, the $\mathcal{S}^\star$-operation on three fields in (4.1) is converted into the total symmetrization on all the four fields at the lagrangian level in (4.2).

Meanwhile the supersymmetry transformation rule (3.7) is also modified, as

$$\delta'_Q \lambda' = +\tfrac{1}{8} \gamma^{\mu\nu} \epsilon F'_{\mu\nu}$$
$$+ \alpha^2 \mathcal{S}^\star \Big[ -\tfrac{1}{128}(c_1 + 4c_3 - 2)(F' \star F')_\nu{}^\nu \star F'_{\rho\sigma}\gamma^{\rho\sigma}\epsilon + \tfrac{1}{16}(c_3 - 1)(F' \star F' \star F')_{\mu\nu}\gamma^{\mu\nu}\epsilon$$
$$- \tfrac{1}{384}(3c_3 + 1) F'_{\mu\nu} \star F'_{\rho\sigma} \star F'_{\tau\lambda} \gamma^{\mu\nu\rho\sigma\tau\lambda}\epsilon \Big] + \mathcal{O}(\alpha^2\lambda^3) \ , \tag{4.3a}$$

$$\delta'_Q A'_\mu = -(\overline{\epsilon}\gamma_\mu \lambda')$$
$$+ \alpha^2 \mathcal{S}^\star \Big[ -\tfrac{1}{16}(c_1 + 2c_2 - 6)(F' \star F')_\nu{}^\nu \star (\overline{\epsilon}\gamma_\mu \lambda') + \tfrac{1}{4}(c_2 - 4)(F' \star F')_\mu{}^\nu (\overline{\epsilon}\gamma_\nu \lambda')$$
$$- \tfrac{1}{8}(-c_2 + 2c_3 + 2) F'_{\mu\nu} \star F'_{\rho\sigma} \star (\overline{\epsilon}\gamma^{\nu\rho\sigma}\lambda')$$
$$- \tfrac{1}{16}(c_2 + c_3 - 1) F'_{\rho\sigma} \star F'_{\lambda\tau} \star (\overline{\epsilon}\gamma_\mu{}^{\rho\sigma\lambda\tau}\lambda') \Big] + \mathcal{O}(\alpha^2 F\lambda^2) \ , \tag{4.3b}$$

up to $\mathcal{O}(\alpha^2\lambda^3)$ or $\mathcal{O}(\alpha^2 F\lambda^2)$ as in (3.7). Since not only the fields $A_\mu$ and $\lambda$ but also the supersymmetry transformation rule itself is changed, we need to put the prime also on $\delta_Q$ itself. To be more specific, the modification in $\delta_Q A_\mu$ is understood as

$$\Delta(\delta_Q A_\mu) \equiv \delta'_Q A'_\mu - \delta_Q A_\mu \equiv (\delta'_Q - \delta_Q) A'_\mu + \delta_Q(A'_\mu - A_\mu) \ , \tag{4.4}$$

where the first term is the modification of the transformation rule, interpreted as

$$(\delta'_Q - \delta_Q) A'_\mu = [\,-(\overline{\epsilon}\gamma_\mu \lambda')\,] - [\,-(\overline{\epsilon}\gamma_\mu \lambda)\,] \ . \tag{4.5}$$

These results (4.2) through (4.5) are in agreement with [8][10], despite the noncommutativity inherent in our system.

## 5. Concluding Remarks

In this paper, we have presented the noncommutative version of supersymmetric non-Abelian gauge theory in 10D given by the action $I_{\text{NCSYM}}$ (2.1). We have seen that despite the noncommutativity yielding a potentially non-vanishing term (2.12) in the invariance check under (2.4), it actually vanishes, thanks to the $\gamma$-matrix algebra (2.13), as well as the basic property (2.6) of noncommutativity.

We have next presented an action $I_{\text{NCSDBI}}$ with higher-derivative terms (3.2) up to quintic terms, as a noncommutative generalization of a supersymmetric DBI action in 10D, which is also regarded as a supersymmetrization of noncommutative a DBI action. With the frequent aid of basic equations of noncommutative geometry, in addition to the usage of the $\mathcal{S}^\star$-operator for the total symmetrization of $\star$ products, we have found that all the $\mathcal{O}(\alpha^2)$ terms with $\star$ products of four fields cancel each other, in the supersymmetric variation of our action $I_{\text{NCSDBI}}$ under supersymmetry (3.7).

We have also clarified possible ambiguities of coefficients in certain terms in the action, in terms of field redefinitions of the $A_\mu$ and $\lambda$-fields. This situation is completely the same as the commutative case [8][10], even with the exact matching of coefficients.

Our result leads to the next natural trial of all the possible dimensional reductions into dimensions lower than 10D, acquiring all the known and possibly unknown noncommutative gauge theories in these lower dimensions, including those in 4D. To put it differently, we can take the advantage of high dimensions as in 10D, *via* dimensional reductions that generate more possibilities compared with the direct formulations in 4D.

Once the most fundamental case of $U(N)$ Yang-Mills group has been established for noncommutative supersymmetric DBI action, it is much easier to apply the recent technique of anti-automorphism of $\star$-matrix algebra [7], in order to get other gauge algebras, *e.g.*, the $o_\star(N)$ or $usp_\star(N)$ algebra, with more phenomenological applications.

Our result also indicates that there is no fundamental obstruction for constructing noncommutative supersymmetric DBI action presumably in any space-time dimensions. As we have seen, the system somehow arranges itself, and automatically avoids any new problem caused by the noncommutativity, as long as we use the total symmetrization operator $\mathcal{S}^\star$, even in such subtle computations with higher-derivatives at $\mathcal{O}(\alpha^2)$. This result also strongly suggests that the possibility of noncommutative supergravity that has never been established in the past, even though there seems some potential problem with defining spinors in space-time with complex metric. Our recent formulation of noncommutative gravity based on teleparallelism [11] might well be a good starting point for such a purpose.

Once our noncommutative supersymmetric gauge theory containing the next-leading terms in noncommutative DBI action is established in 10D, all other lower-dimensional descendant theories are generated by dimensional reductions. In this sense, our theory plays a role of the master theory for noncommutative supersymmetric DBI theory.